\documentstyle[eqsecnum,aps,twocolumn]{revtex}
\input{psfig}

\def\Journal#1#2#3#4{{#1} {\bf #2} (#4) #3}

\def\PRL{Phys. Rev. Lett.}
\def\PRB{{Phys. Rev.} B}

\def\be{\begin{equation}}
\def\ee{\end{equation}}
\def\bea{\begin{eqnarray}}
\def\eea{\end{eqnarray}}

\begin{document}
\draft
\preprint{third version, August 11, 1997}
\wideabs{
\title{Inter-layer edge tunneling and transport properties\\
in separately contacted double-layer quantum Hall systems
}
\author{Daijiro Yoshioka}
\address{
Department of Basic Science, University of Tokyo, Komaba\\
3-8-1 Komaba, Meguro-ku, Tokyo 153, Japan
}

\maketitle
\begin{abstract}
A theory of transport in the quantum Hall regime is developed for
separately contacted double-layer electron systems.
Inter-layer tunneling provides a channel
for equilibration of the distribution functions in the two layers
at the edge states.
Resistances and transresistances for various configurations of the electrodes
are calculated as functions of the inter-layer tunneling amplitude.
Induced current in one of the layer by a current in the other
is calculated also.
It is shown that reflection at the leads causes change in the results
for some electrode configurations.
The results obtained in this work is consistent with recent experiments.
\end{abstract}

\pacs{Keywords: edge state, double layer, quantum Hall effect, tunneling\\
}
}

\narrowtext

\section{Introduction}
\label{sec:level1}

Recently it has become possible to fabricate a multi-layer quantum Hall system,
where current and voltage leads are attached separately
to each layers\cite{eisen}.
For such a system it is expected that
the resistances of one of the layer is affected by the presence of the
other, and voltage or current is induced in the other layer
due to the inter-layer interactions.
One of the interaction relevant to the present situation is the Coulomb
interaction between the layers.
When current flows in one of the layer, the interaction induces voltage
in the other layer, the phenomenon known as the Coulomb
drag, and measured as transresistance\cite{gramila,rubel}.
Another important interaction is the electron tunneling between the
layers.
Actually, in a recent experiment by Ohno {\it et al}. where tunneling is
allowed\cite{ohno},
the Hall and
the longitudinal resistances on one layer is affected considerably
by the presence of the other.
While the effect of the Coulomb drag is relatively weak and expected to
vanish at zero temperature,
the effect of tunneling is larger in the typical situation, and
persists even to zero temperature.
We have established a formulation to calculate resistances
for various situations in the presence of the tunneling\cite{yoshi,yoshi1}.
We have obtained results which successfully explain the experimental results.
In the calculation the effect of the Coulomb drag is neglected, since it
is small.
The results are independent of temperature as far as it is low enough that the
integer quantum Hall effect is observable.

In this paper we extend our theory to include the effect of contact resistance
at the lead.
It is taken into account as transmission and reflection probabilities
between the leads and the edge states of the two-dimensional electrons.
In the experiments leads are attached to one of the layers only,
which is accomplished by blocking
the contact to the other layer by a gate.
We notice that this blocking can be described as perfect reflection at the
leads.

In Section 2 we introduce our theory briefly.
Resistances defined for the layer where the current mainly
flows are calculated in Section 3.
We show that in some cases negative resistance is realized.
In Section 4 the transresistances are shown to have simple relations
to the ordinary resistances.
Current induced in one layer when a current flows in the other is
calculated in Section 5.
Discussion is given in Section 6.
%
%
\begin{figure}
\psfig{figure=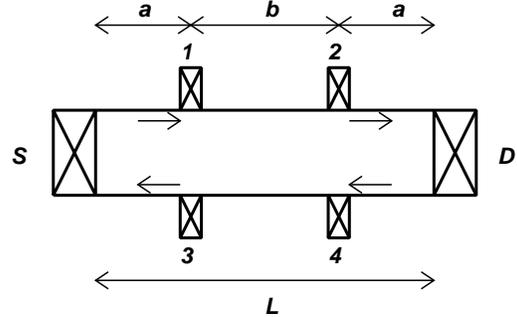,height=5cm}
\caption{
Geometry of the sample.
Two identical layers are stacked vertically.  The leads
(source, drain, 1, 2, 3, and 4) can contact both layers or either layer.
The drift directions along the edges are indicated by arrows.
The lengths $L$, $a$, and $b$ specify the length of the sample and
the positions of the leads.
\label{fig1}
}
\end{figure}
%
%
\section{formulation}
\label{sec:level2}

We consider a Hall bar type sample shown in Fig.1.
The energy levels in the bulk are adjusted to be aligned in the two layers,
and we consider the system in the lowest quantum Hall regime.
For a single layer system the
quantum Hall effect can be understood as a consequence of the spatial
separation between left-going and right-going states on opposite edges  of the
sample\cite{butt}.
At each edge we can define the chemical potential, which stays constant
between the leads, and which differs between the opposite edges.
On the other hand, the situation is quite different for
a double-layer system with finite inter-layer tunneling.
Here the electron distribution in the edge states may not be
in the thermal equilibrium.
Therefore at first sight the chemical potential seems difficult to
be defined.
However, we can still define effective local chemical potential
at each position of the edges of each layer\cite{yoshi}.
This introduction of the chemical potential opened a way to calculate
resistances.
It is a peculiarity of the one-dimensionality and chirality of the
edge state that the current depends only on the effective chemical
potential.
Therefore our result is independent of the temperature.

We take $x$-axis along the edge and the effective chemical potential
is written as $\mu_\sigma (x)$,
where $\sigma = \pm$ represents the layer index.
Then the following equation governs the development of the chemical
potentials between the leads:
\begin{equation}
\frac{d\mu_{\sigma}(x)}{dx} = - \frac{1}{\xi}[\mu_{\sigma}(x)
-\mu_{-\sigma}(x)],
\label{boltchem}
\end{equation}
where $\xi$ is a parameter meaning the relaxation length for interedge
equilibration.
The current along an edge in layer $\sigma$ at position $x$ is given  by
\begin{equation}
I_{\sigma} (x) = \frac{e}{h}[\mu_{\sigma}(x) - \epsilon_{0}],
\label{currentsolv}
\end{equation}
with $\epsilon_{0}$ being a common reference energy.

The voltage leads will not affect the chemical potential
as long as they are attached separately to each layer,
since the current do not flow through the lead.
On the other hand the current leads give discontinuous change
to the chemical potential.
When current $I$ flows into the minus-layer of the
sample at the source, the following equations should be satisfied.
\be
I = \frac{e}{h}[\mu_-(0) - \mu_-(2L)],
\label{current}
\ee
where $2L$ is the circumference of the sample,
therefore $\mu_-(0)$ is the chemical potential of the edge state where
electrons are injected from the source, and $\mu_-(2L)$ is that where
they flow back into the source.
By denoting the transmission and reflection probabilities between
the edge states and the source by $t_{\rm s}$ and $r_{\rm s}$
($=1-t_{\rm s}$), respectively,
the chemical potential of the
source, $\mu_{\rm s}$, is given as follows:
\be
\mu_-(0) = t_{\rm s} \mu_{\rm s} + r_{\rm s} \mu_-(2L),
\label{musource}
\ee
and the current can be written as
\be
I = \frac{e}{h}t_{\rm s} [\mu_{\rm s} - \mu_-(2L)].
\label{iandmu}
\ee
In the experiments metallic gates are placed near the leads,
and adjusted to disconnect one of the layer from the lead.
It corresponds to making the transmission probability zero
for the layer to be disconnected.
Thus it is plausible that the transmission probability to the other
layer is reduced.
This is one of the reason we consider non-ideality of the leads here.

One of the equations Eq.(\ref{current}) - Eq.(\ref{iandmu})
gives the boundary condition for Eq.(\ref{boltchem}).
Solutions for various situations are given in the next section.

\section{Resistances}
\label{sec:level3}

Here we consider two experimental situations both of which are
actually realized experimentally.
First we consider the simplest case where only one of the layers,
the minus layer,
has connection to source and drain where current flows in and flows out.
The voltages are measured in the same layer by leads 1 to 4 (see Fig.1),
and resistances are obtained.
(If the chemical potential is measured in the plus layer, it gives
transresistance.
We will discuss it in the next section.)
In this case the resistances are independent of the transmission
probability provided that it is finite.
With the distance between the leads given by $a$ and $b$ as shown in
Fig.1, and by the notation
$\alpha = \exp(-2L/\xi)$,
the longitudinal resistance $R_{xx} =R_{12} = R_{34}$ and the Hall resistance
$R_{xy} = R_{13}$ are given as follows:
\begin{equation}
R_{xx} = \frac{h}{e^{2}}
\frac{\alpha^{a/L}(1 - \alpha^{b/L})}{2(1+\alpha)},
\label{rxx}
\end{equation}
\begin{equation}
R_{xy} = \frac{h}{e^{2}}
\frac{1+\alpha^{a/L}+\alpha^{1-a/L} + \alpha}
{2(1+\alpha)}.
\label{rxy}
\end{equation}
On the other hand the two-point resistance defined by
$R_{\rm sd}=
(\mu_{\rm s} - \mu_{\rm d})/eI$ depends on $t_{\rm s}$ and $t_{\rm d}$:
\be
R_{\rm sd}=
\frac{h}{e^2}[\frac{1}{t_{\rm d}}
+ \frac{1}{t_{\rm s}} - \frac{1+3\alpha}{2(1+\alpha)}]
\label{twoR},
\ee
where $t_{\rm d}$ is the transmission probability at the drain.
The dependence of $R_{xx}$ and $R_{xy}$ on $L/\xi$ is shown in Fig.2.
The Hall resistance is quantized only in the strong or weak coupling limit,
where $L/\xi = \infty$, $\alpha=0$, or $L/\xi =0$, $\alpha=1$, respectively.
%
%
\begin{figure}
\psfig{figure=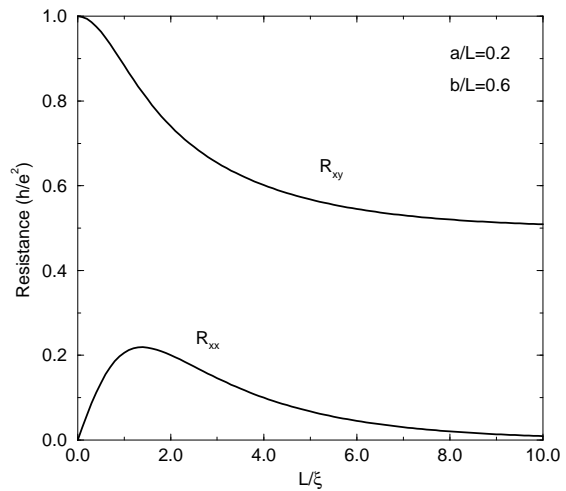,height=7cm}
\caption{
Longitudinal and Hall resistances as a function of $L/\xi$.
Here only the minus layer is connected to the leads.
\label{fig2}
}
\end{figure}
%
%

In another situation we consider here, the two layers are both connected
to the common source, while only minus layer is connected to the drain.
In this case the resistances depend on the transmission and reflection
probabilities at the source and drain.
The general expression is too complicated to be written here.
The analytical expressions for the perfect
leads case, $t_{\rm s}=t_{\rm d}=1$, is
\be
R_{12}=0, ~~R_{34} = \frac{h}{2e^2}\alpha^{a/L}(1-\alpha^{b/L}),
\label{case21}
\ee
and
\be
R_{13}=\frac{h}{2e^2}(1+\alpha^{(a+b)/L}).
\label{case22}
\ee
These results and the case of non-ideal leads where
$t_{\rm s}=t_{\rm d}=0.2$ are shown in Fig.3.
The asymmetry of the longitudinal resistance at the upper and lower edges of
the sample comes from the chirality of the edge current in the magnetic field.
They are interchanged by the reversal of the magnetic field
direction\cite{note}.
It should be noticed that the longitudinal resistance $R_{12}$
becomes negative for non-ideal case.
Therefore the edge current flows adversely in the chemical potential
gradient.
%
%
\begin{figure}
\psfig{figure=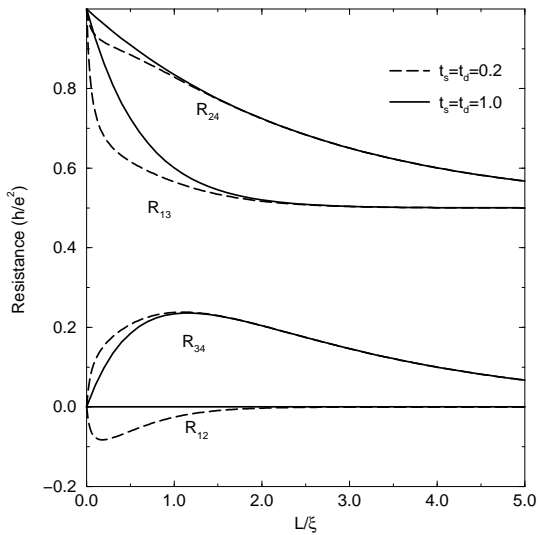,height=3.0in}
\caption{
Longitudinal and Hall resistances as a function of $L/\xi$.
Here both layers are connected to the source.
The solid lines are for the ideal lead case and dashed lines
for non-ideal leads, where $t_{\rm s}=t_{\rm d}=0.2$ and $r_{\rm s}=r_{\rm d}=0.8$.
\label{fig3}
}
\end{figure}
%
%

\section{Transresistances}
\label{sec:level4}

In our system most of the current flows in the minus layer.
The voltages measured in the plus layer give transresistances.
They are measured especially for the experiments of the Coulomb drag.
In the present paper, Coulomb interaction is neglected, but still the
transresistances are finite due to the tunneling.

In our system the two edge states at the opposite side of the sample are not
coupled directly.
Therefore even though the current in each layer is not conserved, the sum of
the currents of the two layers at each side of the sample $I_+(x) + I_-(x)$
does conserve.
Thus $\mu_+(x) + \mu_-(x)$
is also a constant along any edge between the current leads.
Then the following relations for the transresistance, $R_{ij}^{\rm t}$,
is easily verified:
\be
R_{ij} + R_{ij}^{\rm t} = 0,
\label{trans1}
\ee
if $i$ and $j$ are on the same side, and
\be
R_{ij} + R_{ij}^{\rm t} = \frac{h}{e^2},
\label{trans2}
\ee
if $i$ and $j$ are on the opposite side.
The transresistances are easily obtained from  the
results of the previous section.

\section{Induced current}
\label{sec:level5}

When any two leads in the plus layer is connected by a conducting wire
in the presence of the current flow in the minus layer,
some current generally flows the wire in the absence of direct external
connection.
We assume that the resistance of the wire is negligible.
The contact resistance at the leads, given by the transmission and reflection
probabilities, are considered as before, however.

%
%
\begin{figure}
\psfig{figure=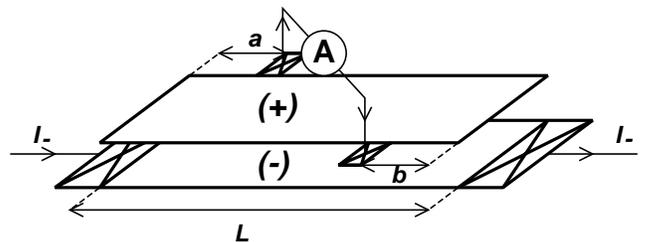,height=4cm}
\caption{
The configuration of the leads in the plus layer where current is
measured.  The positive direction of the current is indicated by
arrows in the figure.
\label{fig4}
}
\end{figure}
\begin{figure}
\psfig{figure=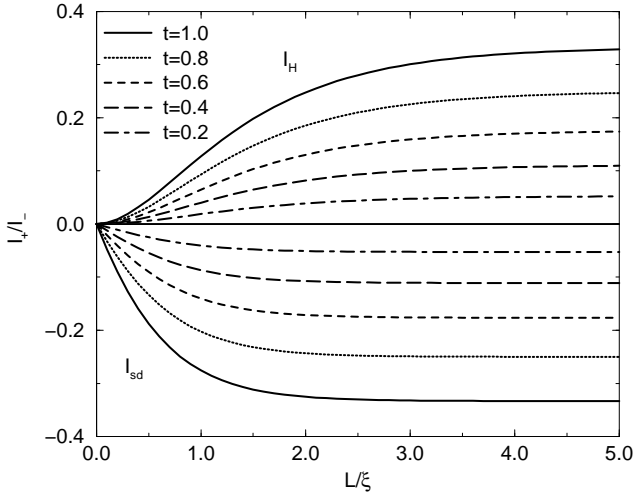,height=7cm}
\caption{
Induced current $I_+$ in the plus layer by a current in the minus layer,
$I_-$, where
$t$ is the trassmission probability at the leads.
When the source and drain are connected ($a=b=0$ in Fig.4),
$I_+ \equiv I_{\rm sd}$ is negative.
While when the edges are connected in the middle ($a=b=L/2$ in Fig.4),
$I_+ \equiv I_{\rm H}$ is positive.
\label{fig5}
}
\end{figure}

%
%
The analytical expression
for the general case can be obtained,
but the expressions are too complicated to be written here.
Thus we give here the results of two simple cases:
i) the source and drain of the plus layer is connected and ii) the leads at
the middle of the sample are connected.  See Fig.~\ref{fig4}:
the case i) corresponds to $a=b=0$ and the case ii) to $a=b=L/2$.
For ideal leads analytical expression for case i) is
\begin{equation}
\frac{I_{\rm sd}}{I_{-}} = \frac{\alpha -1}{\alpha + 3},
\end{equation}
and for case ii) it is
\begin{equation}
\frac{I_{\rm H}}{I_-} = \frac{(1-\sqrt{\alpha})^2}{\alpha + 3}.
\end{equation}
In these equations $\alpha =\exp(-2L/\xi)$ as before, and
$I_-$ is the external current through the source and drain in the minus layer.
The numerical results including the non-ideal cases are given
in Fig.~\ref{fig5},
where $t$ is the transmission probability at the leads.
As expected the contact resistance reduces the induced current monotonously.
%
%
%
%

\section{Discussion}
\label{sec:level6}

In recent experiments Ohno {\it et al.}~\cite{ohno}
observed $R_{xy} \simeq h/e^2$ in the
presence of finite $R_{xx}$ in the situation considered in Section 2.
This result is at a first glance inconsistent with our result, since
$L/\xi \simeq 50$ and $\alpha \simeq 0$,
if estimated using $\Delta_{\rm SAS} = 0.02$ meV\cite{yoshi}.
However, the inconsistency is resolved, if the edges are not aligned
perfectly.
In this case the tunneling amplitude is reduced considerably.
In the experiments they observed that
the effect of in-plane magnetic field is quite small.
This insensitivity to the tilting is also in accordance with
non-ideal alignment as discussed in Ref.\cite{yoshi1}.
They have realized the situation of Fig.3 also.
The behavior at such a situation is also consistent with the present
theory.

In the experiment by Ohno {\it et al.} the coupling between
the two layers are already
weak so that it cannot be reduced further by the in-plane magnetic
field.
However, if the sample used has large tunneling probability between the
edge states, the in-plane magnetic field should have noticeable effect,
especially if the field is tilted perpendicularly to the edge.
In this case the system should move from strong to
weak tunneling regime
as we increase the in-plane field.
In the course, the longitudinal resistance should first increase and
after showing a peak decreases again.
As noticed in the previous work\cite{yoshi}, the height of the peak is
in some sense universal: it only depends on the ratio
between $a$, $b$, and $L$, and independent of the actual size of the
sample.
At the same time, in the situation of Fig.~\ref{fig3}, the negative
longitudinal resistance should be observed.
We hope such experiments will be done in the near future.

\acknowledgments

Part of this work is done at Indiana University in collaboration with
A.H. MacDonald and also at Aspen Center for Physics.
The author thanks hospitality of IU and ACP.


\begin{references}
\bibitem{eisen} J.P. Eisenstein, L.N. Pfeiffer, and K.W. West,
\Journal{Appl. Phys. Lett.}{57}{2324}{1990}.

\bibitem{gramila} T.J. Gramila, J.P. Eisenstein, A.H. MacDonald,
L.N. Pfeiffer, and K.W. West, \Journal{\PRL}{66}{1991}{1216}.

\bibitem{rubel} H. Rubel, A. Fischer, W. Dietsche, K. von Klitzing,
and K. Eberl, \Journal{\PRL}{78}{1763}{1997}.

\bibitem{ohno} Y. Ohno, M. Foley, and H. Sakaki,
\Journal{\PRB}{54}{R2319}{1996}.

\bibitem{yoshi} D. Yoshioka and A.H. MacDonald, \Journal{\PRB}{53}
{R16168}{1996}.

\bibitem{yoshi1} D. Yoshioka and A.H. MacDonald,
in High Magnetic Fields in the Physics of Semiconductors II,
ed. G. Landwehr and W. Ossau (World Scientific, Singapore)
pp. 207-210.

\bibitem{butt} M. B\"uttiker, \Journal{\PRB}{38}{9375}{1988}.

\bibitem{note} The role of the source and drain is also interchangeable.
Our results are independent of the sign of $\mu_{\rm s}-\mu_{\rm d}$.
Electron can be injected from the drain by making $\mu_{\rm s}<\mu_{\rm d}$.

\end{references}
\end{document}